\documentclass[aps,pra,epsfigure,twocolumn,showpacs]{revtex4}
\usepackage{dcolumn}
\usepackage{bm}
\usepackage{graphicx}
\usepackage{amsmath}
\usepackage{latexsym}
\usepackage{amsfonts}
\usepackage{amssymb}
\usepackage{array}
\usepackage{epsfig}
\usepackage{times}

\newcommand{\ket}[1]{\left\vert#1\right\rangle}
\newcommand{\miniket}[1]{\vert#1\rangle}

\newcommand{\bra}[1]{\left\langle#1\right\vert}
\newcommand{\minibra}[1]{\langle#1\vert}
\newcommand{\minisand}[3]{\langle#1\vert#2\vert#3\rangle}

\newcommand{\nbar}{\overline{n}}

\begin{document}
\title{Information-flux approach to multiple-spin dynamics}
\author{C. Di Franco $^1$,  M. Paternostro$^1$, G. M. Palma$^2$, and M. S. Kim$^1$}
\affiliation{$^1$School of Mathematics and Physics, Queen's University, Belfast BT7 1NN, United Kingdom\\
$^2$NEST-CNR (INFM) \& {D}ipartimento di Scienze Fisiche ed Astronomiche, Universita' degli studi di Palermo, via Archirafi 36, 90123, Italy}

\begin{abstract}
We introduce and formalize the concept of information flux in a many-body register as the influence that the dynamics of a specific element receive from any other element of the register. By quantifying the information flux in a protocol,  we can design the most appropriate initial state of the system and, noticeably, the distribution of coupling strengths among the parts of the register itself. The intuitive nature of this tool and its flexibility, which allow for easily manageable numerical approaches when analytic expressions are not straightforward, are greatly useful in interacting many-body systems such as quantum spin chains. We illustrate the use of this concept in quantum cloning and quantum state transfer and we also sketch its extension to non-unitary dynamics.
\end{abstract}

\maketitle
 
\section{Introduction} 
\label{Intro}

Very recently, we have witnessed a strong interest of the quantum information processing (QIP) community in problems of interacting many-body systems~\cite{generale}. Such interest is motivated by the unresolved necessity of tackling, theoretically and experimentally, quantum processors equipped with large registers. The investigation of multipartite devices has been boosted by the observation that specific forms of built-in and permanent intra-register couplings can be used for the purposes of quantum communication and quantum computation~\cite{generale,alwayson}. Indeed, it has been shown that the control over such systems can be sensibly reduced in a way to avoid the generally demanding fast and accurate inter-qubit switching and gating (we will name such a scenario as {\it control-limited}). However, the price to pay for the performance of efficient operations is the pre-engineering of appropriate patterns of couplings. Determining the exact distribution of coupling strengths for a given inter-qubit interaction model in a control-limited setting is often a matter of craftsmanship or the result of the exploitation of peculiar topological properties of the physical system at hand~\cite{cambridge}. A more systematic approach to such a problem is evidently in order. 

Here, we provide a tool for this task by introducing the concept of {\it information flux} in a quantum mechanical system. An intuitive understanding of our idea can be given in terms of the influences that the dynamics of a selected element of a multipartite register experience due to {\it interaction channels} with the other parties. We show that a desired protocol is performed by arranging the network of interactions in a way to privilege or repress specific interaction channels. The realization of an optimal QIP task is therefore translated into the maximization of the information flux associated with such channels. This result is twofold interesting: conceptually, the study of information flux in multipartite schemes for QIP allows for a deeper insight into the physical mechanisms behind an investigated protocol. We provide a significant example, in this paper, by addressing the case of universal quantum cloning machines where the achievement of the optimal cloning fidelity is shown to correspond to an information-flux maximization problem. Pragmatically, as anticipated, it helps in designing the most appropriate pattern of interaction strengths and/or initial state of a register. We address the case of information transfer in spin chains as a scenario where such a possibility is particularly useful. Although our attention is focused on registers of qubits, it is important to notice that the dimensionality of the Hilbert space of the elements of a register does not represent a limitation to the effectiveness of our approach. We also sketch the way interaction channels can be generalized to the non-unitary case by explicitly introducing noise and decoherence in the system and considering the corresponding open-dynamics information flux.

This paper is organized as follows. In Sec.~\ref{concept} we introduce the approach we use to analyze different schemes for QIP. In Sec.~\ref{UQCM} we investigate the quantum circuit proposed by B\v{u}zek {\it et al.}~\cite{buzek} for an optimal two-copy universal quantum cloner (UQCM $1\rightarrow{2}$). An implementation of this protocol, by means of spin chains, is studied in Sec.~\ref{UQCMchain}, where a clear picture of the advantages offered by the information-flux approach is provided. We consider quantum state transfer across a spin chain in Sec.~\ref{transfer}, where we also show how to adapt our approach to handy numerical analysis which is used to reveal interesting features of long and disordered chains. In Sec.~\ref{opensystem} we extend the information-flux concept to non-unitary dynamics. Finally, in Sec.~\ref{remarks} we summarize our results.

\section{Concept}
\label{concept}
Let us consider a register of $N$ interacting qubits coupled via the Hamiltonian $\hat{\cal H}_{\{g\}}(t)$ whose structure we do not need to specify here. Our assumption is that $\hat{\cal H}_{\{g\}}(t)$ depends on a set of parameters $g_j$ (which could stand for the coupling strengths between the elements of the register) and a generalized time parameter $t$. We adopt the notation according to which $\hat{\Sigma}_j=\otimes^{j-1}_{k=1}\hat{\openone}_k\otimes\hat{\sigma}_{\Sigma_{j}}\otimes^{N}_{l=j+1}\hat{\openone}_{l}$ $(\Sigma=X,Y,Z,I)$ is the operator that applies the $\hat{\sigma}_{\Sigma}$ Pauli-matrix only to the $j$-th qubit of the register. Here, $\hat{\sigma}_{I_j}\equiv\hat{\openone}_j$ with $\hat{\openone}_{j}$ the $2\times{2}$ identity matrix of qubit $j$. In the remainder of this paper, unless explicitly specified, we work in the Heisenberg picture where time-evolved operators are indicated as $\hat{\tilde\Sigma}_j(t)=\hat{\cal U}^\dag\hat{\Sigma}_j\hat{\cal U}$ with $\hat{\cal U}(t)={\rm exp}[{-({i}/{\hbar})\int\hat{\cal H}_{\{g\}}(t')dt'}]$. 
\begin{figure}[t]
\psfig{figure=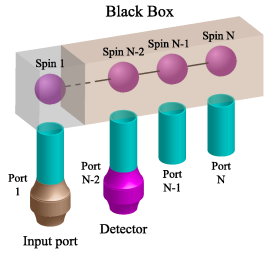,width=3.4cm,height=3.3cm}
\caption{(Color online) Scheme of the processes we consider. A computation or communication step is interpreted as a black box (whose operation depends on the coupling scheme within a multipartite register of qubits) with movable input and a detection terminal. Through the study of information-flux dynamics, we can design the best coupling scheme for a chosen QIP operation.}
\label{fig:circ4}
\end{figure}
We say that {\it there is information \it extractable from qubit $j$ at time $t$} whenever there is at least one $\Sigma$ (obviously excluding the identity) for which $\langle\hat{\tilde\Sigma}_{j}(t)\rangle\neq{0}$. Here, the expectation value is calculated over the initial state of the register $\ket{\Psi_0}_{1..N}$. In some cases, the analysis of expectation values in the Heisenberg picture can be considerably helpful in gathering information about the dynamics of a system~\cite{gottesman}.

We adopt, hereafter, the following schematic description of a computation or communication process: we suppose to have access to a selected qubit of a multipartite register and we consider it as the input terminal of the black box given by the rest of the elements. We then use a detection stage which can be attached to a suitable output port, connected to one of the qubits in the black box (a sketch is given in Fig.~\ref{fig:circ4}). In this picture, the initial state of a quantum system is described by the state vector $\ket{\Psi_0}_{1..N} = \ket{\phi_0}_1 \ket{\psi_0}_{2...N}$. This is the case in which the first qubit is initialized in a generic {\it input} state (and is separable with respect to the rest of the register) and $\ket{\psi_0}_{2...N}$ represents the initial state of the other qubits that, in general, can be mutually entangled. We assume this state to be known and independent of the input state. The assumption of a known initial state of the register $\{2,..,N\}$ is physically motivated as it corresponds to the situation assumed in many control-limited QIP protocols~\cite{generale,alwayson,cambridge,spincloning}. We can thus interpret a quantum process as the {\it flux} of appropriately processed information from the input qubit to the remaining components of the register. Such a flux is witnessed by any explicit dependence of the dynamics of the $i$-th qubit from the operators associated with the input one. Therefore, in order to find if qubit $i$ has developed any extractable information at time $t$, as a result of an information flux from the input qubit, 
we need to study the dependence of $\minisand{\Psi_0}{\hat{\tilde\Sigma}_i(t)}{\Psi_0}$'s on at least one of $\minisand{\Psi_0}{\hat{\tilde\Sigma}_1(0)}{\Psi_0}$'s. In what follows, we show that the study of the dynamics of these expectation values, taking advantage from the possibility of working with ``known and fixed'' states, allows for proper interaction engineering and optimization that can be used for variegated tasks of control-limited QIP. 

Each $\hat{\tilde{\Sigma}}_i(t)$ can be decomposed over the operator-basis built out of all the possible tensorial products of single-qubit operators acting on the elements of the system $\{1,..,N\}$. Such a basis has dimension $4^N$ and its elements can be partitioned in four disjoint groups, each of dimension $4^{N-1}$. Each group contains operators having the form $\hat{\sigma}_{\Sigma_1}\otimes\hat{\cal G}_{k,2..N}$ ($k=1,..,4^{N-1}$) with $\hat{\cal G}_{k,2..N}$ being the tensorial product of single-qubit operators acting on the elements of the system $\{2,..,N\}$. By using the knowledge we have about the initial state of the $\{2,..,N\}$ register, we introduce the vector ${\bm \chi}$ such that $\chi_k=\phantom{}_{2..N}\!\bra{\psi_0}\hat{\cal G}_{k,2..N}\ket{\psi_0}_{2..N}$. Therefore
\begin{equation}
\label{IFformale}
\minisand{\Psi_0}{\hat{\tilde\Sigma}_i(t)}{\Psi_0}=\sum_{\Sigma'=X,Y,Z,I}\sum^{4^{N-1}}_{k=1}\alpha^{\Sigma\Sigma'}_{ik}(t)\chi_{k}\minisand{\phi_0}{\hat{\sigma}_{\Sigma'_1}}{\phi_0}.
\end{equation}
Here, the matrix ${\bm \alpha}^{\Sigma\Sigma'}(t)$ incorporates the details of the time-evolved multi-qubit operators and is determined once $\hat{\cal H}_{\{g\}}(t)$ is assigned. The coefficient ${\cal I}^{\Sigma\Sigma'}_{i}(t)=\sum^{4^{N-1}}_{k=1}\alpha^{\Sigma\Sigma'}_{ik}(t)\chi_{k}$ defines and quantifies the flux of extractable information (or, shortly, the {\it information flux}) from $\hat\Sigma'_1$ to $\hat{\Sigma}_i$ at time $t$. The formal definition of the information flux highlights the dual nature of the control we can operate over the dynamics of a multi-qubit system. Indeed, besides the dynamical part of ${\cal I}^{\Sigma\Sigma'}_{i}(t)$, there is a time-independent part that incorporates information about the initial state of the register, input qubit apart. This represents an additional control over the dynamics at hand. Information flux towards a specific element of the register can be depleted or enhanced by properly preparing the register $\{2,..,N\}$ in a way to engineer $\chi_{k}$'s. Notice that the intrinsic dependence of the Hamiltonian on the set $\{g\}$ makes the information flux implicitly dependent on the coupling scheme being chosen. Obviously, by choosing a different partition for the four groups of operators in which the basis for an $N$-qubit system has been divided, Eq.~(\ref{IFformale}) can be straightforwardly modified in order to make explicit which is the information flux from $\Sigma_{j}$ to $\Sigma'_{i}$. In our black-box model, this corresponds to a change in the position of the input terminal (cf. Fig.~\ref{fig:circ4}). 
\begin{figure}[b]
\hskip1.5cm{\bf (a)}\hskip4.1cm{\bf (b)}\\
\vskip0.15cm
\psfig{figure=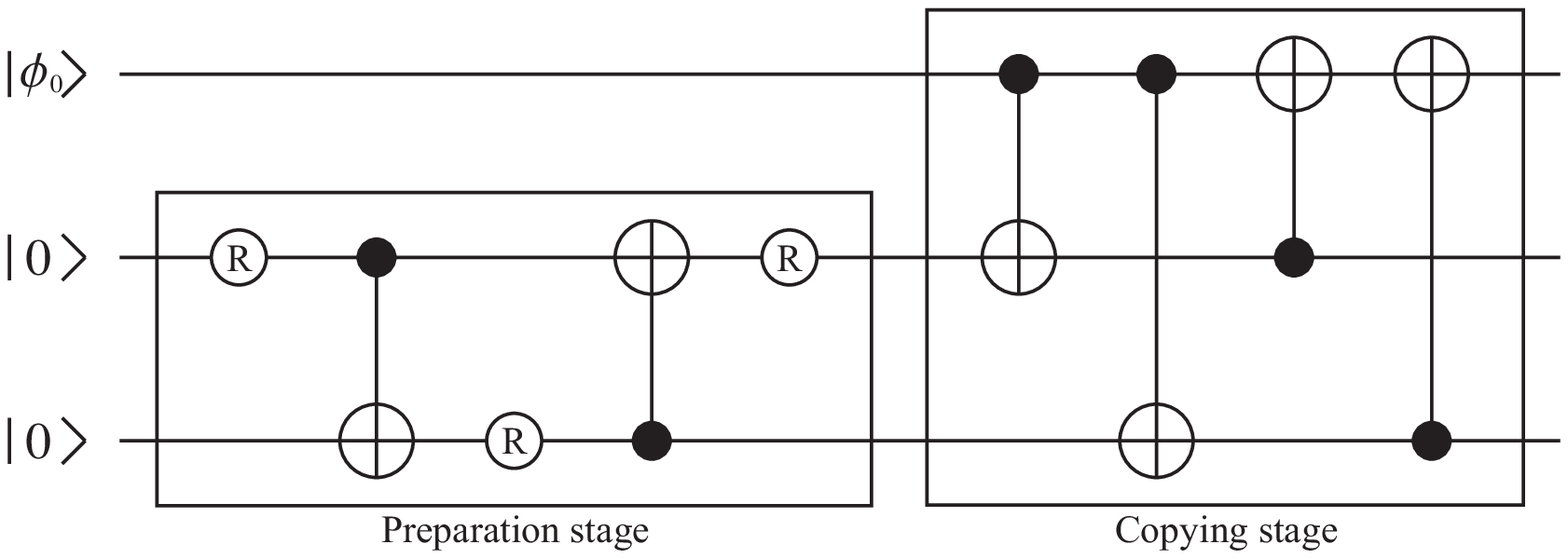,width=5.5cm,height=2cm}\hskip0.5cm\psfig{figure=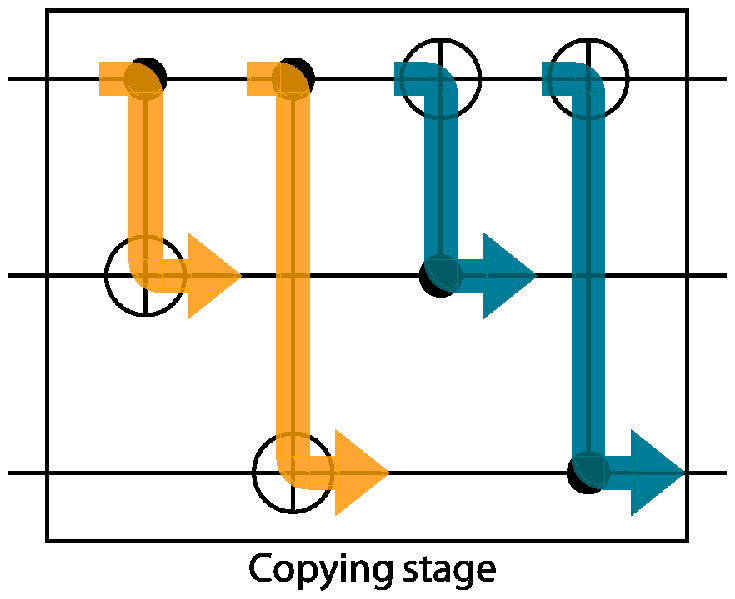,width=2.5cm,height=2cm}
\caption{(Color online) {\bf (a)}: Logical circuit for optimal UQCM $1 \rightarrow 2$~\cite{buzek}. Each line represents a qubit and time flows from left to right. We show single-qubit rotations, indicated by the letter $R$ inside a circle, and two-qubit CNOT gates. {\bf (b)}: Information flux from the input qubit to the rest of the register in the copying stage of the UQCM $1 \rightarrow 2$. The yellow arrows represent the $\sigma_z$ information fluxes and the blue arrows represent the $\sigma_x$ information fluxes.}
\label{fig:circ1}
\end{figure}
The dual control discussed above can be fully utilized in the preparation of a multipartite device in the most appropriate configuration (of couplings and initial state) for a given QIP task. The potentialities of this approach are better illustrated by means of explicit examples. In what follows, we address the important problems of quantum cloning and quantum information transfer. The use of the information flux approach in the investigation of the latter problem, in particular, turns out to be quite useful for the analysis and characterization of quantum state transfer in a long spin chain with imperfections.

\section{Universal Quantum Cloning machine $1 \rightarrow 2$}
\label{UQCM}
We consider the problem of an optimal two-copy universal quantum copy machine (UQCM $1 \rightarrow 2$), for which extensive literature is available~\cite{cloning}. In our study we concentrate on the quantum circuit proposed by B\v{u}zek {\it et al.}~\cite{buzek} and shown in Fig.~\ref{fig:circ1} {\bf (a)}. In order to adapt the framework of Ref.~\cite{buzek} to the notation used so far, we say that qubits $2$ and $3$ embody the register described in the previous section and qubit $1$ is the input qubit. The circuit is composed of two different stages: the {\it preparation} stage, composed of single-qubit rotations $R$'s and controlled-NOT (CNOT) gates, prepares the second and the third qubit in the state $\ket{\psi_0}_{23} =(1/\sqrt{6}) ( 2 \ket{00}_{23} + \ket{01}_{23} + \ket{10}_{23})$ while the state to be cloned is encoded in the input qubit. The {\it copying} stage in Fig.~\ref{fig:circ1} {\bf (a)} produces two optimal clones of the input state on the second and the third qubit. The cloning fidelity is $\frac{5}{6}$, which is the maximum for a UQCM $1 \rightarrow 2$~\cite{bruss}. 

Let us now use an information-flux approach to the problem by studying the flux from $\Sigma_{1}$ to $\Sigma'_{2,3}$. Of course we need to investigate only the copying stage, because no interaction involves the first qubit during the preparation stage, which can be performed off-line. The copying stage is built out of four CNOT gates, which altogether give the time-evolution operator. We only need to consider $\hat{X}_{2,3}(t)$ and $\hat{Z}_{2,3}(t)$, for instance, as $\hat{Y}_{2,3}(t)$ is found as $i\hat{X}_{2,3}(t)\hat{Z}_{2,3}(t)$. For a general CNOT gate acting on its control and target qubits~\cite{nielsenchuang}, it is immediate to get the following relations:
\begin{equation}
\hat{\tilde{X}}_c= \hat{X}_c\hat{X}_t,\hskip0.42cm\hat{\tilde{X}}_t =\hat{X}_t,\hskip0.42cm\hat{\tilde{Z}}_c=\hat{Z}_c,\hskip0.42cm\hat{\tilde{Z}}_t=\hat{Z}_c\hat{Z}_t,
\end{equation}
with $\hat{X}_c$, $\hat{Z}_c$ ($\hat{X}_t$, $\hat{Z}_t$) the operators of the control (target) qubit before the action of the gate and $\hat{\tilde{X}}_c$, $\hat{\tilde{Z}}_c$ ($\hat{\tilde{X}}_t$, $\hat{\tilde{Z}}_t$) the analogous operators after the action of the gate. Obviously, in this example, the generalized ``time-parameter'' corresponds to the ``logical application of a single CNOT gate.'' By using these results, we straightforwardly find the expressions of the evolved $\hat{X}_{i}$ and $\hat{Z}_{i}$ ($i=1,2,3$) for the register of the UQCM $1\rightarrow{2}$ after each step of the copying stage. These are summarized in Table~\ref{tab1}~\cite{commento1}.
\begin{table}[b]
\begin{ruledtabular}
\label{tabella}
\begin{tabular}{c c c c c} \hline
    & $t_1$ & $t_2$ & $t_3$ & $t_4$ \\ \hline
$\hat{\tilde{X}}_1 (t)$ & $\hat{X}_1\hat{X}_2$ & $\hat{X}_1\hat{X}_2\hat{X}_3$ & $\hat{X}_1\hat{X}_2\hat{X}_3$ & $\hat{X}_1\hat{X}_2\hat{X}_3$ \\
$\hat{\tilde{Z}}_1(t)$ & $\hat{Z}_1$ & $\hat{Z}_1$ & $\hat{Z}_2$ & $\hat{Z}_1\hat{Z}_2\hat{Z}_3$\\
$\hat{\tilde{X}}_2(t)$ & $\hat{X}_2$ & $\hat{X}_2$ & $\hat{X}_1\hat{X}_3$ & $\hat{X}_1\hat{X}_3$ \\
$\hat{\tilde{Z}}_2(t)$ & $\hat{Z}_1\hat{Z}_2$ & $\hat{Z}_1\hat{Z}_2$ & $\hat{Z}_1\hat{Z}_2$ & $\hat{Z}_1\hat{Z}_2$ \\
$\hat{\tilde{X}}_3(t)$ & $\hat{X}_3$ & $\hat{X}_3$ & $\hat{X}_3$ & $\hat{X}_1\hat{X}_2$ \\
$\hat{\tilde{Z}}_3(t)$ & $\hat{Z}_3$ & $\hat{Z}_1\hat{Z}_3$ & $\hat{Z}_1\hat{Z}_3$ & $\hat{Z}_1\hat{Z}_3$ \\ \hline
\end{tabular}
\end{ruledtabular}
\caption{The evolved operators after time $t_j$ [{\it i.e.}, after the application of the $j$-th CNOT in Fig.~\ref{fig:circ1} {\bf (a)}] expressed in terms of the operators before the copying stage.}
\label{tab1}
\end{table}
Our first observation is that there is no information flux from $\hat{X}_1$ to $\hat{X}_{2,3}$ during the first interaction, as $\hat{\tilde{X}}_{2,3} (t_1)$ do not depend on $\hat{X}_1$. The same holds for $\hat{\tilde{Z}}_{3}(t_1)$.
On the other hand, by applying the formalism introduced in the previous section, an easy calculation let us see that 
$\minisand{\Psi_0}{\tilde{Z}_2 (t_1)}{\Psi_0} = \minibra{\phi_0}\minisand{\psi_0}{Z_1 Z_2}{\psi_0}\miniket{\phi_0} = \frac{2}{3} \minisand{\phi_0}{Z_1}{\phi_0}$, {\it i.e.}, ${\cal I}^{ZZ}_2(t_1)=2/3$. After $t_1$, no information flows from $\hat{Z}_1$ to $\hat{Z_{2}}$, as it can be seen from Table~\ref{tab1}, so that ${\cal I}^{ZZ}_2(t_4)={\cal I}^{ZZ}_2(t_1)$. An analogous reasoning leads us to conclude that the remaining non-zero information fluxes in this problem are given by ${\cal I}^{XX}_{2,3} (t_4)={\cal I}^{YY}_{2,3} (t_4)={\cal I}^{ZZ}_3(t_4)=2/3$. In this case, where all ${\cal I}^{\Sigma\Sigma}_{2,3}$'s are equal and ${\cal I}^{\Sigma\Sigma'}_{2,3}$'s vanish for $\Sigma\ne\Sigma'$, we can affirm that the Bloch vectors of the output states are proportional to the one associated with the input state with the {\it shrinking factor} being equal to the information flux. The timing and pattern of information fluxes are pictorially illustrated in Fig.~\ref{fig:circ1} {\bf (b)}.
The specific pattern of fluxes is a particular result of the symmetry of the UQCM $1 \rightarrow 2$.
In general, when the flux is restricted to homonymous operators, the following simple relation can be derived connecting the {\it fidelity} ${\cal F}_{1i}(t)$ between the input state and the state of qubit $i$ at time $t$ and the corresponding information fluxes in the register 
\begin{equation}
\label{fidelity}
{\cal F}_{1i}(t)=\frac{1}{2}\left[1+\!\!\sum_{\Sigma=X,Y,Z}{\cal I}^{\Sigma\Sigma}_{i}(t)(_1\bra{\phi_0}{\hat{\sigma}_{\Sigma_1}}\ket{\phi_0}_1)^2\right].
\end{equation}
In the specific case at hand and regardless of the input state, ${\cal F}_{12}(t_4)={\cal F}_{13}(t_4)=5/6$, as it should be. Equation~(\ref{fidelity}) can be straightforwardly generalized to the case where cross-operator fluxes are present. Such an expression is of no relevance, however, for the purposes of this paper.

The pedagogical example discussed above, however, does not exhaust the possibilities offered by an information-flux approach. Indeed, as anticipated, the method can also be used to find the optimal preparation stage of a circuit designed to perform a given task. If we consider the above-mentioned copying stage, without knowledge of the input state, we can use the information-flux analysis to design the proper preparation for an optimal UQCM $1 \rightarrow 2$. Let us suppose that, starting only from the information-flux analysis previously done on the copying stage, we want to deduce the initial state of the second and the third qubit to obtain an optimal UQCM $1 \rightarrow 2$. The conditions on such an initial state can be easily obtained by imposing that the two clones are identical ({\it symmetry condition}) and the cloning process is independent of the input state ({\it universality condition}). The symmetry condition implies that ${\cal I}^{\Sigma\Sigma}_{2}(t_4)={\cal I}^{\Sigma\Sigma}_{3}(t_4)$ while the universality condition gives ${\cal I}^{\Sigma\Sigma}_{2,3}(t_4)={\cal I}^{\Sigma'\Sigma'}_{2,3}(t_4)$ (with $\Sigma,\Sigma'=X,Y,Z$). From the study of the copying stage we know that no cross-operator flux is possible~\cite{commento1}, so that the use of Eq.~(\ref{fidelity}) is legitimate. For the generic input $\ket{\phi_0}_1=c_0\ket{0}_{1}+c_1\ket{1}_{1}$ ($|c_0|^2+|c_1|^2=1$), we have that $(\bra{\phi_0}\hat{\sigma}_{X_1}\ket{\phi_0})^2+(\bra{\phi_0}\hat{\sigma}_{Y_1}\ket{\phi_0})^2+(\bra{\phi_0}\hat{\sigma}_{Z_1}\ket{\phi_0})^2=1$. As the process must be optimal, the fidelities have to be maximized (with the constraints imposed on the information fluxes), regardless of the state to be cloned. This means the maximization of the information flux. For a generic input state $\ket{\psi_0}_{23} = \alpha \ket{00}_{23} + \beta \ket{01}_{23} + \gamma \ket{10}_{23} + \delta \ket{11}_{23}$, this requirement implies the maximization of $\minisand{\psi_0}{X_2}{\psi_0} = \minisand{\psi_0}{Z_2}{\psi_0} = \minisand{\psi_0}{X_3}{\psi_0} = \minisand{\psi_0}{Z_3}{\psi_0}$ which results in $\alpha = \sqrt{{2}/{3}}$, $\beta = \gamma={1}/{\sqrt{6}}$, and $\delta = 0$. We therefore retrieve, through an information-flux analysis, the initial state of the optimal UQCM $1\rightarrow{2}$.

As an example of how the initial preparation of the register can help in designing the optimal operation to perform by implicitly ``inhibiting'' information flux, let us consider the following simple but yet illustrative case. Suppose that, using again the copying stage of Figs.~\ref{fig:circ1}, we want to obtain a circuit that generates the output state of qubit $2$ equal to the input state, regardless of the third qubit (in some sort of fully-biased asymmetric cloning). Again, we want to design an appropriate preparation stage for qubits $2$ and $3$. The complete asymmetry of the process requires (quite intuitively) that ${\cal I}^{\Sigma\Sigma}_{2}(t_4)=1$. By using Table~\ref{tab1} again, this turns out to be equivalent to $\minisand{\psi_0}{X_3}{\psi_0} =  \minisand{\psi_0}{Z_2}{\psi_0} = 1$. For the generic input state $\ket{\psi_0}_{23}$ used above, these conditions are satisfied only for $\alpha =\beta={1}/{\sqrt{2}}$ and $\gamma=\delta=0$, so that $\ket{\psi_0}_{23} =({1}/{\sqrt{2}})\ket{0}_{2} ( \ket{0}_{3} + \ket{1}_{3} )$. The preparation stage should thus be a simple Hadamard transform over the third qubit. 

\section{UQCM $1 \rightarrow 2$ with spin chains}
\label{UQCMchain}
The information-flux approach offers remarkable advantages also in the cases where, rather than abstract quantum circuits, we have to deal with physical interaction models described by coupling Hamiltonians. This situation is particularly relevant for scenarios of control-limited QIP, where multi-qubit interactions are often employed for the engineering of appropriate quantum interference processes resulting in the performance of QIP tasks~\cite{generale,cambridge,spincloning,spinoptimalcloning,mauro}. In particular, it has been realized that information transfer and quantum cloning can be performed through well-known multi-qubit models such as $XY$ or Heisenberg couplings~\cite{generale,spincloning}. 
Here, we focus our attention to the optimal UQCM $1 \rightarrow 2$ realized with a three-qubit chain~\cite{spinoptimalcloning} whose coupling Hamiltonian, within a nearest-neighbor anisotropic Heisenberg model, reads $\hat{\cal H} = ({J}/{2})\sum_{i=1}^2 (\hat{X}_i\hat{X}_{i + 1} +\hat{Y}_i\hat{Y}_{i + 1}+ \lambda\hat{Z}_i\hat{Z}_{i + 1})$. Here, $J$ is the coupling strength and $\lambda\ge{0}$ is the anisotropy parameter. The central qubit (labeled by 2) is the input one, encoding the state to be cloned, while the external qubits $1$ and $3$ are the output ones. Note that this is a situation requiring the input terminal in our black-box schematics to be shifted to qubit $2$. In an information-flux perspective we will thus be interested in the fluxes from qubit $2$, this time. Here, we {\it demonstrate}, by means of the proposed new tools, that for $\lambda=2$, a proper initialization of qubits $1$ and $3$ and a careful timing allow for optimal UQCM $1\rightarrow{2}$~\cite{spinoptimalcloning}.

The symmetry condition, in this case, is a direct consequence of the physical system at hand so that, in evaluating the information flux from qubit 2 during the evolution, we can just concentrate on qubit $1$. From the analysis conducted so far on UQCM $1\rightarrow{2}$, it appears that the conditions of no cross-flux and ${\cal I}^{\Sigma\Sigma}_{2,3}(t^*)=2/3$ ($\forall~\Sigma=X,Y,Z$ and at a proper time $t^*$) guarantee the universality condition and the optimal value of fidelity to be achieved. A detailed calculation shows that the information fluxes between homonymous operators of qubits $2$ and $1$ (or equivalently $3$) depend on $_{13}\!\bra{\psi_0}\hat{\Sigma}_{1}\hat{\Sigma}_{3}\ket{\psi_0}_{13}$ with $\ket{\psi}_{13}$ a generic two-qubit pure state. In order for such expectation values to be physically meaningful when we impose ${\cal I}^{\Sigma\Sigma}_{2,3}(t^*)=2/3$ ({\it i.e.} in order to have $_{13}\!\bra{\psi_0}\hat{\Sigma}_{1}\hat{\Sigma}_{3}\ket{\psi_0}_{13}\in[-1,1]$), the only possible choice for the anisotropy parameter is $\lambda=2$~\cite{commentocinesi}. In turn, it implies that ${}_{13}\!\bra{\psi_0}\hat{X}_{1}\hat{X}_{3}\ket{\psi_0}_{13}=_{~13}\!\bra{\psi_0}\hat{Y}_{1}\hat{Y}_{3}\ket{\psi_0}_{13}=1$ and $_{13}\!\bra{\psi_0}\hat{Z}_{1}\hat{Z}_{3}\ket{\psi_0}_{13}=-1$. The only state of qubits $1$ and $3$ satisfying these conditions is $\ket{\psi_+}_{13}=(1/\sqrt 2)(\ket{01}+\ket{10})_{13}$, so that $\ket{\psi_{0}}_{13}=\ket{\psi_+}_{13}$. With these choices, ${\cal I}^{\Sigma\Sigma}_{2,3} (t)=({2}/{3})\sin^2 (\sqrt{3} J t)$ and at $t^*={\pi}/({2 \sqrt{3} J})$ we have the optimal fidelity ${\cal F}_{21}(t^*)={\cal F}_{23}(t^*)=5/6$. Our rigorous result confirms the ``guess'' for the initial state of the clones in Ref.~\cite{spinoptimalcloning} and demonstrates the uniqueness of such a preparation for the optimality of the process.

There is a profound difference between the cloning performed through the multi-spin coupling and the formal circuital dynamics depicted in Fig.~\ref{fig:circ1} {\bf (a)}, which the information-flux approach is able to clearly spot out. For cloning realized with the Heisenberg coupling, the information fluxes from the input to the output qubits are always equal during the evolution. As shown in Fig.~\ref{fig:circ1} {\bf (b)}, on the other hand, the quantum circuit allows the information flow in the register in a sequential manner. Therefore, while the universality condition is always respected during the UQCM $1 \rightarrow 2$ realized with a multi-spin coupling, this condition holds in the quantum circuit only considering the whole transformation, but not during the single steps performed by the quantum gates. The same holds for the symmetry condition. Therefore, if we stop the interaction in the spin system at a time different from $t^*$, the cloning fidelity will certainly be lower than the optimum but it will not depend on the input state and it will be the same for both the clones. Our method can also be applied to the very recently proposed case of UQCM $1\rightarrow{N}$ through spin-chain models~\cite{cinesinuovi}.
 
\section{State transfer}
\label{transfer}
An important part of our study is the investigation of quantum state transfer (QST) from the information-flux viewpoint. We show how such a change of perspective with respect to the standard approach to QST~\cite{generale,cambridge,mauro,burgarth}, allows for the study of long-chain dynamics in the presence of imperfections in a convenient way. Let us gradually arrive to such a result.

It is well-known that perfect QST is possible for a three-qubit chain of equal $XY$ couplings governed by ${\cal \hat{H}} = ({J}/{2}) \sum_{i=1}^2(\hat{X}_i \hat{X}_{i + 1} +\hat{Y}_i \hat{Y}_{i + 1})$~\cite{cambridge}. This can be seen by explicit calculation of the spectrum of $\hat{\cal H}$ and the corresponding fidelity. In our analysis, the same result is achieved simply by noticing that, in these conditions, ${\cal I}^{XX}_{3}(t) = {\cal I}^{YY}_{3}(t) = \sqrt{{\cal I}^{ZZ}_{3}(t)}=\sin^2 ({J t}/{\sqrt{2}})$, so that at $t^*= \pi / \sqrt{2} J$ there is a unit information-flux from each operator of the first qubit to the homonymous operators of the last one. Incidentally, we notice that an imprecision in $t^*$ affects ${\cal I}^{ZZ}_{3}(t)$ much more than the other fluxes. This means that the transfer fidelity in a process which is stopped at the wrong interaction time drops faster for input states with a large modulus of the $\sigma_z$-expectation value. This explains the frailness of the QST process with input states having a large population-inversion.

The achievement of perfect QST by using a long linear chain of qubits all prepared in their ground state is known to be possible for weighted coupling strengths following the pattern $J_i = \lambda \sqrt{i(N-i)}$ with $i=1,..,N-1$ labeling a site in the chain: After a time $t^* = \frac{\pi}{\lambda}$, we obtain perfect state transfer from the first to the last qubit~\cite{cambridge}. While this result can be retrieved through the information-flux approach, here we would like to be more pragmatic. The achievement of a unit transfer fidelity, although obviously desirable, might not be necessarily in order. Depending on the specific protocol that has to be realized after the QST process, a lower threshold to ${\cal F}_{1N}$ may be enough (if ${\cal F}_{1N}$ is sufficiently large, for instance, one can use state-purification procedures after the receipt of the state). This can be helpful in practical QST implementations if, by paying the price of a lower transfer fidelity, the demanding requirements over the coupling pattern are relaxed. Here we show that this possibility is actually realistic.

For some specific coupling configurations and assuming periodic boundary conditions, the $XY$ Hamiltonian can be diagonalized by means of a sequence comprising Wigner-Jordan, Fourier and Bogoliubov transformations~\cite{lieb}. However, this does not hold for any pattern of coupling rates. As here we consider an inhomogeneous set of $J_i$'s in an open chain ruled by the Hamiltonian ${\cal \hat{H}} = \frac{1}{2} \sum_{i=1}^{N-1} J_i (\hat{X}_i \hat{X}_{i + 1} +\hat{Y}_i \hat{Y}_{i + 1})$, we decide not to rely on these methods~\cite{reuter}. By means of a simple application of the operator-expansion formula, on the other hand, it is possible to design an efficient numerical apparatus to analyze the information flux. The method is based on the construction of {\it recurrence formulas} building up the flux of information between operators of specific qubits along the chain. For instance, for the non-trivial case of $N=5$ we have 
\begin{equation}
\begin{aligned}
\hat{\tilde{X}}_5 (t)&= a(t) \hat{X}_5 + b(t) \hat{Z}_5 \hat{Y}_4 + c(t) \hat{Z}_5 \hat{Z}_4 \hat{X}_3\\
& + d(t) \hat{Z}_5 \hat{Z}_4 \hat{Z}_3 \hat{Y}_2+ e(t) \hat{Z}_5 \hat{Z}_4 \hat{Z}_3 \hat{Z}_2 \hat{X}_1. 
\end{aligned}
\end{equation}
The time-dependent coefficient $a(t)$ can be obtained by the formula $a(t) = \sum_{i = 0}^{\infty} a_{2i} (-1)^i \frac{(2t)^{2i}}{(2i)!},$ where $a_i$ is related to $b_i$, $c_i$, $d_i$, and $e_i$ through 
\begin{equation}
 \begin{split}
  & a_i = J_4 b_{i-1},~~b_i = J_4 a_{i-1} + J_3 c_{i-1},\\
  & c_i = J_3 b_{i-1} + J_2 d_{i-1},~~d_i = J_2 c_{i-1} + J_1 e_{i-1},\\
  & e_i = J_1 d_{i-1}
 \end{split}
\end{equation}
with the initial conditions $a_0=1,\,b_0=c_0=d_0=e_0=0$. The other coefficients  $b(t)$, $c(t)$, $d(t)$ and $e(t)$ are similarly defined. As the initial state of the chain is $\ket{\Psi_0} = \ket{\phi_0}_1 \ket{0000}_{2..5}$, we have ${\cal I}^{XX}_{5}(t) = e(t)$.

\begin{figure}[t]
{\bf (a)}\hskip4cm{\bf (b)}
\centerline{\psfig{figure=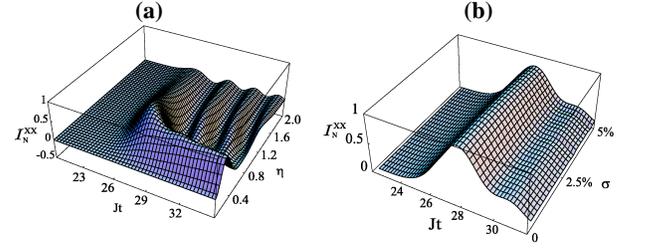,width=8cm,height=2.95cm}}
\caption{(Color online) {\bf (a)}: ${\cal I}^{XX}_N$ in a spin chain of length $N=101$ with equal central coupling strengths, against the rescaled interaction time $J t$ and $\eta$. {\bf (b)}: ${\cal I}^{XX}_N$ for the same conditions in panel {\bf (a)} and $\eta=\eta_{max}$, plotted against $J t$ and the standard deviation $\sigma$ of the Gaussian distribution which determines the disorder parameters $\delta_i$.}
\label{fig:plot2}
\end{figure}

\begin{figure}[b]
\centerline{{\bf (a)}\hskip3cm{\bf (b)}\hskip3cm{\bf (c)}}
\centerline{\psfig{figure=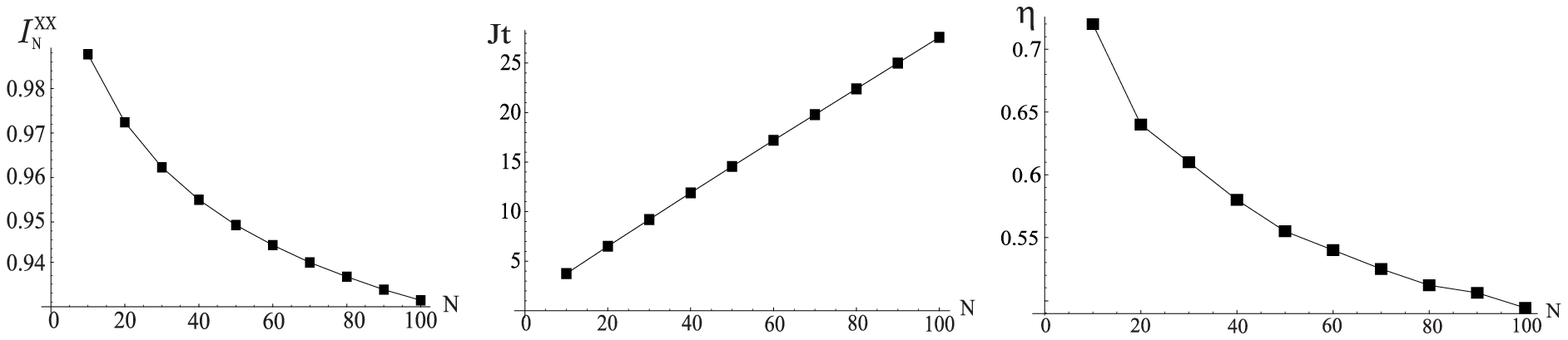,width=9.0cm,height=2.3cm}} 
\caption{[{\bf (a)}, {\bf (b)}, {\bf (c)]} Maximum value of the information flux ${\cal I}^{XX}_N$, the corresponding rescaled time $J t$ and inhomogeneity parameter $\eta$ against the number of spins $N$ for chains of up to $N=101$ qubits.}
\label{fig:plot3}
\end{figure}

Having illustrated the formal technique, we now move to the case of an arbitrary number of qubits. From now on, our {\it limited-control} assumption is that we are able to engineer the strength of the coupling rates of just the extremal qubits ($J_1$ and $J_{N-1}$)~\cite{extremalcouplings}. We take $J_i = J$ (for ${i}=2,.,N-2$), $J_1 = J_{N-1} = \eta{J}$ and study the behavior of the flux ${\cal I}^{\Sigma\Sigma}_N$ against the dimensionless interaction time $J t$ (within a reasonably long range) and the inhomogeneity parameter $\eta$. It is important to stress that the limitations to our analysis are simply the results of the work conditions we have assumed. In principle, any pattern of coupling strengths can be introduced and studied through our approach.
In Fig.~\ref{fig:plot2} {\bf (a)} we show ${\cal I}^{XX}_N$ for the case of $N=101$. Evidently, there are values of $(\eta,Jt)$ for which the information flux towards the last qubit in the chain is almost ideal. For instance, at $\eta\simeq{0.5}$ and $Jt\simeq{27.6}$, ${\cal I}^{XX}_{101}\simeq0.93$. An explicit evaluation reveals that 
${\cal I}^{YY}_N={\cal I}^{XX}_N$ while ${\cal I}^{ZZ}_N$, being equal to their product, reaches its maximum value for the same $\eta$ and rescaled time as ${\cal I}^{XX}_N$. This implies that a transfer fidelity $\ge{0.865}$ is achieved for the worst-case scenario given by the fully polarized state $\ket{1}_1$ being transferred across the chain (here we have used the generalized formula for fidelity, due to the fact that cross-operator fluxes are present). To give a complete picture of the way the transfer dynamics behaves in such a control-limited scenario, in Figs.~\ref{fig:plot3} {\bf (a)}, {\bf (b)} and {\bf (c)} we present the maximum value of ${\cal I}^{XX}_N$ and the corresponding $J t$ and $\eta$ against the length of the chain up to $N=101$. Compared to the perfect transfer case of $N=3$, which occurs for all-equal coupling rates, a decrease down to $J_1=J_{N-1}=J/2$ is sufficient to achieve high-fidelity QST.

The flexibility of our approach is well-witnessed by the easiness with which imperfections in the coupling rates' pattern can be incorporated. This is an important point in any realistic assessment of QST as it will be difficult, in practice, to arrange an ideal distribution of coupling strengths. We have therefore considered {\it disordered} $J_{i}$'s, each deviating from the value corresponding to the QST described above by a quantity $\delta_{i}$ that follows a Gaussian distribution centered at $0$ and with a standard deviation {$\sigma$} of, at most, $5\%~J_{i}$~\cite{disorder}. For $N=101$ qubits, the corresponding ${\cal I}^{XX}_N$ is shown in Fig.~\ref{fig:plot2} {\bf (b)}. The performance of the chain is quite robust against the disorder. In particular, an interesting feature that can be experimentally helpful is that the time at which the flux is maximized does not change significantly with the strength of the disorder with the maximum being only slightly dependent on it. 

\section{Open-system dynamics}
\label{opensystem}
Here, we sketch how the framework presented in this paper can be extended so as to address the case of open-system dynamics where the register is subject to inter-element interactions and is also exposed to environmental effects. In this case, as we explain, the information flux will contain a purely coherent part, arising from the unitary dynamics of the system at hand, and an incoherent contribution. The starting point of our study is a master equation (ME) for the open dynamics of the register. For the sake of simplicity, we assume a Markovian environment, described by a Liouvillian $\hat{\cal L}(\varrho)$ acting on $\varrho$, the density matrix of the system. To fix the ideas, we consider the case where each qubit interacts with its own environment. The generalization of this situation is discussed later in this section. We thus have
\begin{equation}
\label{me}
\begin{split}
\hat{\cal L}(\varrho)&=\sum_{i}\left\{-\frac{\Gamma}{2}(\hat{\sigma}_{-,i}\,\varrho\,\hat{\sigma}_{+,i}-\{\hat{\sigma}_{+,i}\hat{\sigma}_{-,i},\varrho\})\right.\\
&\left.-\frac{\gamma}{2}[\hat{\sigma}_{z,i},[\hat{\sigma}_{z,i},\varrho]]\right\}.
\end{split}
\end{equation}
Here, $\hat{\sigma}_{+,i}$ ($\hat{\sigma}_{-,i}$) is the raising (lowering) operator of qubit $i$, $\Gamma$ is the rate of dissipation of each qubits immersed in a zero-temperature dissipative environment ({\it i.e.}, a thermal bath of mean occupation number $\nbar=0$). Finally $\gamma$ is a dephasing rate. We assume that the register undergoes a simple free evolution and we move to a proper interaction picture, so that the register's ME is $\partial_{t}\varrho=\hat{\cal L}(\varrho)$.  We multiply this by the time-independent operator $\hat{\Sigma}_{k}$ of qubit $k$ and trace over the register so that the ME can be transformed into the set of Langevin-type equations (one for each $\hat{\Sigma}_{k}$)
\begin{equation}
\label{langevinlike}
\begin{aligned}
\partial_{t}\langle\hat{\tilde \Sigma}_k(t)\rangle&=\sum_{i}\left\{-\frac{\Gamma}{2}(\langle\hat{\tilde\sigma}_{+,i}(t)\hat{\tilde\Sigma}_{k}(t)\hat{\tilde\sigma}_{-,i}(t)\rangle\right.\\
&-\langle\hat{\tilde{\Sigma}}_{k}(t)\hat{\tilde\sigma}_{+,i}(t)\hat{\tilde\sigma}_{-,i}(t)\rangle-\langle\hat{\tilde\sigma}_{+,i}(t)\hat{\tilde\sigma}_{-,i}(t)\hat{\tilde{\Sigma}}_{k}(t)\rangle)\\
&\left.-\gamma(\langle\hat{\tilde{\Sigma}}_{k}(t)\rangle-\langle\hat{\tilde\sigma}_{z,i}(t)\hat{\tilde{\Sigma}}_{k}(t)\hat{\tilde\sigma}_{z,i}(t)\rangle)\right\},
\end{aligned}
\end{equation}
where the expectation values are all calculated over the initial state of the register. Obviously, this expression can be easily generalized to the case of $\nbar\neq{0}$ by introducing an additional term to the Liouvillian and to the case where an arbitrary interaction Hamiltonian $\hat{\cal H}_{\{g\}}(t)$ is considered. The time-evolved operators $\hat{\tilde\sigma}_{\pm,i}(t)$ and $\hat{\tilde\sigma}_{z,i}(t)$ can be expressed, in general, in terms of sums of single-qubit operators, weighted by proper time-dependent functions. We therefore immediately recognize that Eq.~(\ref{langevinlike}) (or its generalization) provides the open-system dynamical equations for the information fluxes within the register. In particular, the case at hand addresses just the incoherent evolution of the information fluxes (equivalent to the noise-part of Langevin-type equations for the expectation values of the operators of the system). The described approach can be applied, in exactly the same manner, to the case where the individual-environment assumption does not hold. By studying the information flux between particular states of a register of qubits {\it collectively} interacting with an environment, we gather information about the rate of creation of environmentally-mediated entanglement and the effect that leakage has on it, for instance~\cite{noidetuning}. Non-Markovian ME, addressing the case of environment with memory, can also be treated with our approach.

The use of expectation values over properly designed initial states, once more, can be considered a useful simplification in the analysis of the dynamics of a register. The application of this framework to pragmatic examples, with particular attention to collective or memory-preserving environments, goes beyond the scope of this work and will be the subject of further investigations. 

\section{Remarks}
\label{remarks}
The possibilities offered by control-limited QIP with multi-qubit interactions require the development of exploitable apparatuses for the design of the coupling distributions and preparation stages appropriate for a set task. This is mandatory in order to achieve the optimal pattern of quantum interferences at the basis of any quantum communication and computation protocol. Here, we have introduced the information flux as a promising tool in this respect. We have demonstrated how it can be used in order to gather important insight in the performance of existing protocols for QIP and for their development towards optimality. Particular attention has been paid to the important problem of optimal cloning in interacting spin systems. On the other hand, its intrinsically operative nature allows for the design of quite manageable numerical recipes, particularly important in the manipulation of very large computational registers. These have been instrumental in the study of a long chain of coupled qubits acting as a support for quantum state transfer. We have shown that a simple coupling configuration can be designed allowing for long-haul communication, resilient to (quite large) local fluctuations in the coupling strengths. Finally, we have sketched the way the concept of information flux can be extended to open-system dynamics by means of Langevin-type equations of motion. We believe our technique has rather promising possibilities of application in problems of many-body physics in the presence of disorder and decoherence~\cite{osborne}.

\section{Acknowledgments}
\label{acknowledgments}
C.D.F. would like to thank S. Pantaleone for discussions. We acknowledge financial support from UK EPSRC and QIP IRC. G.M.P. acknowledges support under PRIN 2006 ``Quantum noise in mesoscopic systems'' and under CORI 2006. M.P. is supported by The Leverhulme Trust (Grant No. ECF/40157).


\begin{thebibliography}{99}
\bibitem{commento1} Notice that the analysis of the copying stage prescinds from the specific form of $\ket{\psi_0}_{2..N}$, which enters the calculations only when expectation values (and therefore information fluxes) are required.
\bibitem{commentocinesi} This result is achieved also considering that $-1\le\sum_{~~{13}}\!\bra{\psi_0}\hat{\Sigma}_{1}\hat{\Sigma}_{3}\ket{\psi_0}_{13}\le1$.
\end{thebibliography}
\end{document}